\documentclass{aa}
\usepackage{txfonts}
\usepackage{graphicx}
\usepackage{epsfig}
\usepackage{natbib}
\bibpunct{(}{)}{;}{a}{}{,}
\errorcontextlines = \maxdimen
\newcommand{\cP}{c\,$P_1^{-1}$}
\mathcode`\;="013B
\newdimen\breedte
\begin{document}

\title{Frequency dependence of the drifting subpulses of PSR~B0031$-$07}

\author{J.M. Smits \inst{1} \and D. Mitra \inst{2,3} \and J. Kuijpers \inst{1}}

\institute{Department of Astrophysics, Radboud University Nijmegen, 
Nijmegen \and Max-Planck-Insitut f\"ur Radioastronomy, Bonn \and
National Center for Radio Astrophysics, Pune}

\offprints{J.M. Smits \email{roysm@astro.kun.nl}}

\date{Received <date> / Accepted <date>}

\abstract{The well known drifter PSR B0031$-$07 is known to exhibit
  drifting subpulses where the spacing between the drift bands ($P_3$)
  shows three distinct modes A, B and C corresponding to 12, 6 and 4
  seconds respectively.  We have investigated periodicities and
  polarisation properties of PSR B0031$-$07 for a sequence of 2700
  single pulses taken simultaneously at 328\,MHz and 4.85\,GHz. We
  found that mode A occurs simultaneously at these frequencies, while
  modes B and C only occur at 328\,MHz. However, when the pulsar is
  emitting in mode B at the lower frequency there is still emission at
  the higher frequency, hinting towards the presence of mode B emission at
  a weaker level. Further, we have established that modes A and B
  are associated with two orthogonal modes of polarisation,
  respectively. Based on these observations, we suggest a geometrical
  model where modes A and B at a given frequency are emitted in two
  concentric rings around the magnetic axis with mode B being nested
  inside mode A.  Further, it is evident that this nested configuration
  is preserved across frequency with the higher frequency arising
  closer to the stellar surface compared to the lower one, consistent
  with the well known radius-to-frequency mapping operating in
  pulsars.
}

\maketitle

\section {Introduction}
The single pulses of a pulsar are known to be composed of several
smaller units of emission called subpulses. These subpulses are often
seen to drift in phase across a sequence of single pulses giving rise
to the well-known phenomenon of `drifting subpulses', discovered in
1968 \citep{Drake68}. The drift pattern is seen to repeat itself after
a given time which is usually denoted by $P_3$.  The phenomenon has
since been detected in many pulsars \citep[e.g.][]{Rankin86} and the
process is believed to carry information on the mechanism leading to
coherent radio emission from pulsars. For example, \citet{Ruderman75}
have suggested a vacuum gap model in which the subpulses correspond to
beams of particles (or sparks) produced in the vacuum gap over the
polar cap and are thought to rotate around the magnetic axis due to
the perpendicular component of the electric field and the magnetic
field ($E\times B$ drift). Measurements of the speed of rotation of
subpulses might therefore give direct information on the electric
field in the vacuum gap. Recently, \citet{Deshpande99} have shown that drifting subpulses
observed in PSR B0943+10 can be interpreted as 20 sparks rotating
around the magnetic axis at a uniform speed.

Some pulsars show clear changes in the vertical spacing between drift
bands ($P_3$): for example PSR B0809+74 shows a changing $P_3$ after
it goes through a null \citep{Leeuwen02}. PSR B0031$-$07 is
particularly interesting because it shows three distinct drift-modes
with different $P_3$, which are all very stable. They are named mode
A, B and C and correspond to a $P_3$ of 12, 6 and 4 seconds (or 13, 7
and 4\,$P_1$), respectively. These values are approximations and from
40,000 pulses observed at 327\,MHz \citet{Vivekanand97} found that
they may not be harmonically related. They also found that at 327\,MHz
the relative occurrence rate of these modes are 15.6\%, 81.8\% and
2.6\%, respectively. Furthermore, the pulses occur in clusters
containing 30 to 100 pulses which follow each other with delays
ranging from fifty to several hundred pulse periods. These clusters
are constituted in one of three ways: a series of A bands followed by
B bands, only B bands, or a series of B bands followed by C bands
\citep{Huguenin70, Wright81.3}. This pulsar also shows a clear
presence of Orthogonally Polarised Modes (OPM) in the integrated
position angle sweep \citep{Manchester75}.  Table~\ref{tab:parameters}
lists some of the known parameters of PSR B0031$-$07. The values for
$\alpha$ and $\beta$ were found by fitting the single vector model
from \citet{Rad69} to the position angle of the dominant polarisation
mode. The $\alpha$ and $\beta$ values from the position angle of the
remaining polarisation mode are the same within errors.

PSR B0031$-$07 has been thoroughly studied at low observing
frequencies \citep{Huguenin70, Krishnamohan80, Wright81, Vivekanand95,
Vivekanand97, Vivekanand99, Joshi00}, but only rarely at an observing
frequency above 1\,GHz \citep{Wright81.3, Kuzmin86,
Izvekova93}. \citet{Wright81.3} have observed PSR B0031$-$07 at
1.62\,GHz and have found the same drift-modes as seen at lower
frequencies. \citet{Kuzmin86} have studied the integrated pulse
profiles of PSR B0031$-$07 at 102.7\,MHz, 4.6\,GHz and
10.7\,GHz. \citet{Izvekova93} have studied the subpulse
characteristics of PSR B0031$-$07 at 62, 102, 406 and
1\,412\,MHz. They found that the switching between the three
drift-modes and the nulls occur simultaneously at all
frequencies.\footnote{However, it is not clear from their paper
whether they have sufficient signal to noise at 1414\,MHz to see
single pulses.} In this paper we study the behaviour of the different
modes of drift in PSR B0031$-$07 in radio observations at both low and
high observing frequencies simultaneously. In
Section~\ref{sec:analysis} we explain how the observations have been
obtained, how the different modes of drift have been determined, and
what further analyses have been carried out. In
Section~\ref{sec:results} we present our results. The discussion
follows in Section~\ref{sec:discussion}. In this last section we
present a geometrical model which describes many of the observed
characteristics of this pulsar.

\begin{table}
  \caption{List of known parameters of PSR B0031$-$07.}
  \begin{tabular}{llc}
  \hline
  \hline
  Parameter & Value & Reference \\
  \hline
  $P_1$ & 0.94295\,s& \cite{Taylor93}\\
  $\dot{P}$ & $4.083\cdot10^{-16}$ & ''\\
  DM & 10.89\,pc\,cm$^{-3}$ & ''\\
  $S_{400}$ & 95\,mJy & ''\\
  $S_{1400}$ & 11\,mJy & ''\\
  $B_{surf}$ & $6.31\cdot10^{11}$\,G & ''\\
  $\dot{E}$ & $1.9\cdot10^{31}$\,erg/s & ''\\
  $\alpha$ & 4.5$\pm1.0\degr$ & \\
  $\beta$ & +4.8$\pm1.0\degr$ & \\
  \hline
  \end{tabular}
  \label{tab:parameters}
\end{table}

\subsection{Definitions}
\begin{figure*}
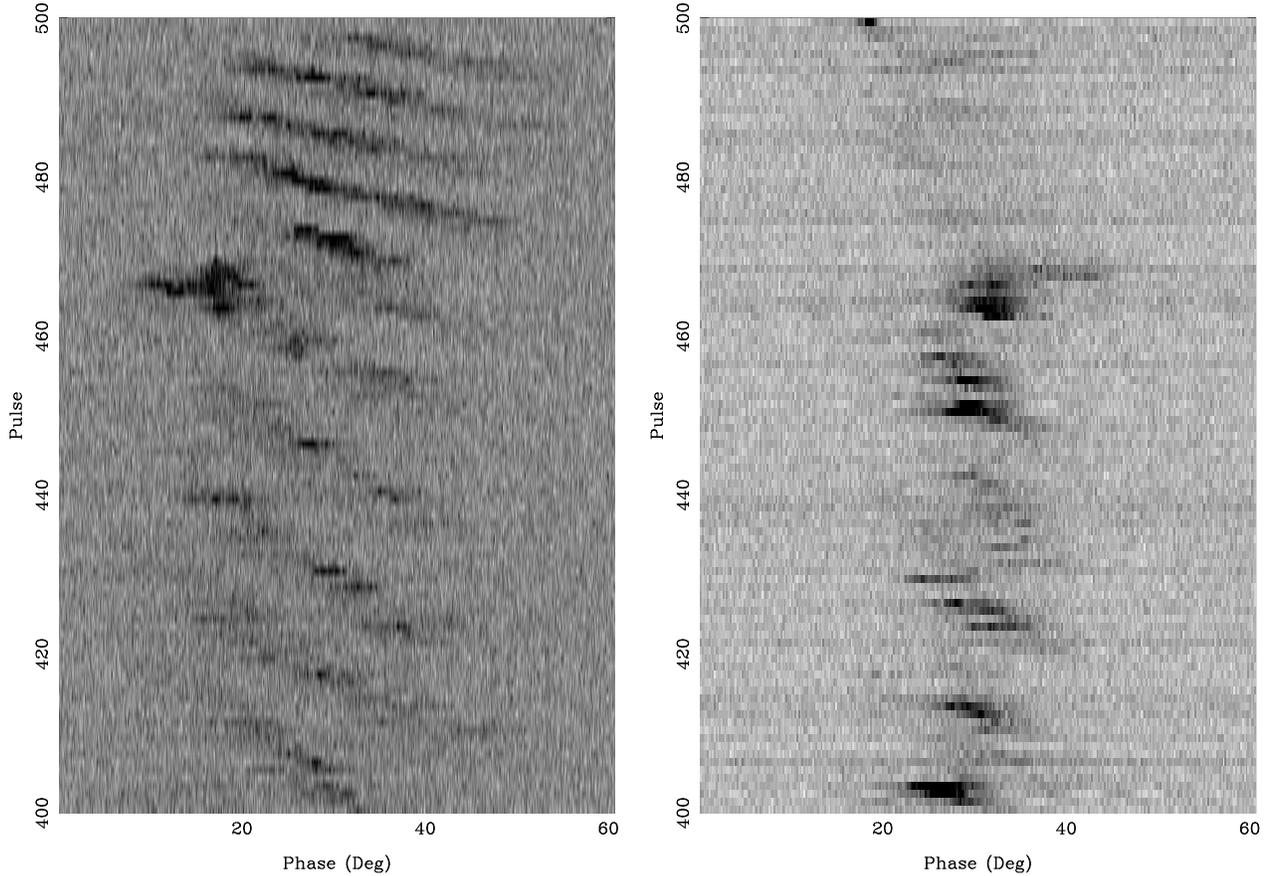

  \centering
  \begin{tabular}{cc}
  \includegraphics[width=0.45\textwidth]{1626f01.eps} &
  \includegraphics[width=0.45\textwidth]{1626f02.eps} \\
  \end{tabular}
  \caption{Gray scale plot of individual pulse profiles from the observation
  at 328\,MHz (left) and 4.85\,GHz (right) (pulse phase
  is plotted horizontally and pulse number vertically). These pulses show an
  example of a rapid mode change. The drift goes from mode A to mode B
  within a few pulses. It can also be seen that there is only a hint
  of mode-B drift in the 4.85-GHz observation.}
  \label{fig:12to6}
\end{figure*}

To describe the observational drift of subpulses we use three
parameters, which are defined as follows: $P_3$ is the spacing at the
same pulse phase between drift bands in units of pulsar periods
($P_1$); this is the ``vertical'' spacing when the individual radio
profiles obtained during one stellar rotation are stacked as in
Fig.~\ref{fig:12to6}. $P_2$ is the interval between successive
subpulses within the same pulse, given in degrees. $\Delta\phi$,
the subpulse phase drift, is the time interval over which a subpulse
drifts, given in $\degr$/$P_1$. Note that $P_2=P_3\times\Delta\phi$.

\section{Data analysis}
\label{sec:analysis}
The observations of PSR B0031$-$07, were obtained on 3 February 2002
with both the Westerbork Synthesis Radio Telescope (WSRT) and the
Effelsberg Radio Telescope simultaneously. These observations were
obtained as part of the MFO\footnote{The MFO collaboration undertakes
simultaneous multi-frequency observations with up to seven telescopes
at any one time.} program. The WSRT observations were made at a
frequency of 328\,MHz and a bandwidth of 10\,MHz. The Effelsberg
observations were made at a frequency of 4.85\,GHz and a bandwidth of
500\,MHz. The time resolutions are 204.8\,$\mu$s and 500\,$\mu$s for
the 328\,MHz and 4.85\,MHz observations, respectively. The 328-MHz
observations have been corrected for Faraday rotation, dispersion and
for an instrumental polarisation effects using a procedure described
in the Appendix of~\citep{Edwards04}. Also, a 50-Hz signal present in
the 4.85-GHz observation has been removed by Fourier transforming the
entire sequence, removing the 50\,Hz peak and Fourier transforming
back. By correlating sequences of single
pulses between the 328-MHz and 4.85-GHz observations that contained
prominent subpulse drift the pulses could be aligned to within an
accuracy of 2$\degr$ of pulse longitude, which confirms the broadband
nature of the pulsar signal. This alignment is sufficient for the
studies presented here.

We have also used an observation from PSR B0031$-$07, obtained on 9 August
1999 with the Effelsberg Radio Telescope at a frequency of 1.41\,GHz
and a bandwidth of 40\,MHz. The time resolution is 250\,$\mu$s.

\subsection{Calculation of $P_3$}
To search for periodicities, we considered a sequence of pulses from
one of the observations. For each pulse in this sequence, we took the flux
at a fixed phase, and calculated the absolute values of the Fourier
transform of this flux distribution. This was done for each phase of
the pulsar window. The resulting transforms were then averaged over
phase, giving a phase-averaged power spectrum (PAPS) from 0 up to 0.5
cycles per rotation period (hereafter \cP), with a
frequency resolution given by the reciprocal of the total length of
the sequence.

Initially, all pulses from the observations were divided into
sequences of 100 pulses, which were searched for peaks in the
PAPS. When a peak was found, the beginning and end of the sequence
was adjusted to get the highest signal-to-noise ratio for the
peak. The signal-to-noise ratio was calculated as the peak value of
the PAPS divided by the rms of the rest of the PAPS. This result was
checked by visual inspection of the sequences to see whether they did
indeed match the beginning and ending of a drift band. $P_3$ was then
calculated as the reciprocal of the centre of the peak in the
PAPS. When a peak would spread over multiple bins, a cubic spline
interpolation was used to determine the location of the peak. The
frequency resolution, given by the number of pulses in the sequence,
was taken as the error on the position of the peak.

Furthermore, we have calculated the PAPS of all pulses of the 1.41-
and 4.85-GHz observations in order to find signs of 6 second
periodicity. For comparison, we also calculated the PAPS of the
328-MHz observation.

\subsection{Values for $P_2$}
Along with values for $P_3$, we also calculated the phase drift for
each sequence of pulses. This was done by cross-correlating consecutive
pulses. The fluxes in an interval around the peak of the cross-correlation were fitted
with a Gaussian, from which the mean was taken as the phase drift. The
error of the fit was taken as the error on the phase drift. $P_2$ was
then determined by multiplying $P_3$ by the phase drift.

\subsection{Average profiles and polarisation properties}
To further study these distinct periodicities, we looked at the
average-pulse profiles for the individual sequences of pulses that show mode A and
mode-B drift, respectively. For the 4.85-GHz observation we compared the
sequences that showed mode-A drift in the 4.85-GHz observation with the
sequences that showed mode-B drift in the 328-MHz observation. In the
same way, we calculated the average linear polarisation, circular
polarisation and position angle as a function of pulse phase. We also
measured the widths of the average total intensity at 10\% and 50\% of the
peak values and at a height three times the rms. These analysis were done
only for the 328-MHz and 4.85-GHz observations.

\section{Results}
\label{sec:results}
\begin{figure*}
  \centering
  \includegraphics[angle=-90, width=0.9\textwidth]{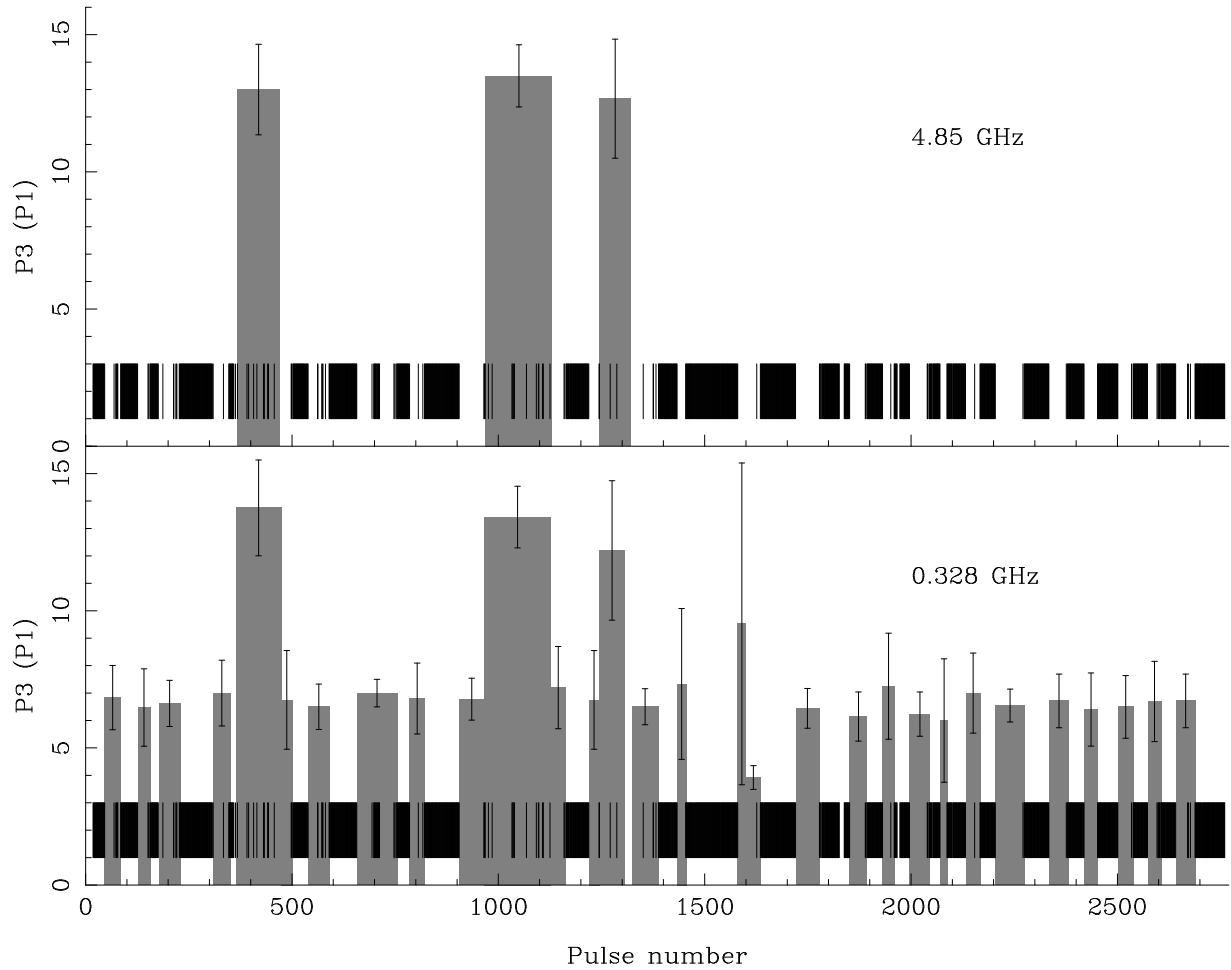} \\
  \caption{Sequences which show a periodicity of magnitude $P_3$ at
    fixed longitude (the gray areas). The lower panel shows the
    328\,MHz observation, while the upper panel shows the 4.85\,GHz
    observation. The black lines indicate pulses which show a null at both
    frequencies.}
  \label{fig:P3}
\end{figure*}

The values of $P_3$ for each sequence and for both frequencies are shown
in Fig.~\ref{fig:P3}. Even though mode B does not seem to occur at
4.85\,GHz, it should be noted that whenever mode B is active at
328\,MHz there is always radiation present at
4.85\,GHz. The same is true for mode C, but there is only one case of
a mode C drift. Fig.~\ref{fig:P3} suggests that the transition between
modes can happen within one or a few pulses. An example of how fast the
drift-rate can change is shown in Fig.~\ref{fig:12to6}. In this figure
the pulses are plotted from bottom to top. Mode A is present in the
first five drift-bands. Then, within one or two pulses, the drift switches
to mode B. In this example, it appears that the transition simply
involves the appearance of a drift-band in a different mode, rather than
the speeding up of the current drift-band. However, it should also be
mentioned that this particular transition happens to occur exactly
when a new drift-band would be expected to arise. In our observations there
are only a few cases when there is a transition from one drift mode into
the other without a null separating the two drift-bands. None of
these transitions show a clear change of mode within one drift-band.

The values for $P_3$ from the observation at 1.41\,GHz, are shown in
Fig.~\ref{fig:P3.1410}. We found various examples of mode-A drift, but
no mode-B drift. In this respect the 1.41-GHz observation resembles
the 4.85-GHz observation. There are no drift-bands showing mode-C
drift.

\begin{figure}
  \centering
  \includegraphics[angle=-90, width=0.45\textwidth]{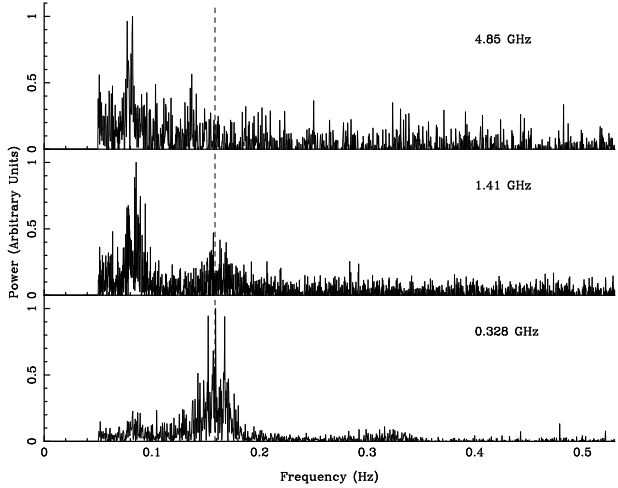}
  \caption{Phase averaged power spectra of three observations at
  frequencies of 4.85\,GHz, 1.41\,GHz and 328\,MHz. There are clear
  signs of 6-seconds periodicity in the 1.41-GHz and 328-MHz observations. The
  dotted line is placed at 1/6.3\,s}
  \label{fig:powers}
\end{figure}

\begin{figure}
  \centering
  \includegraphics[angle=-90, width=0.45\textwidth]{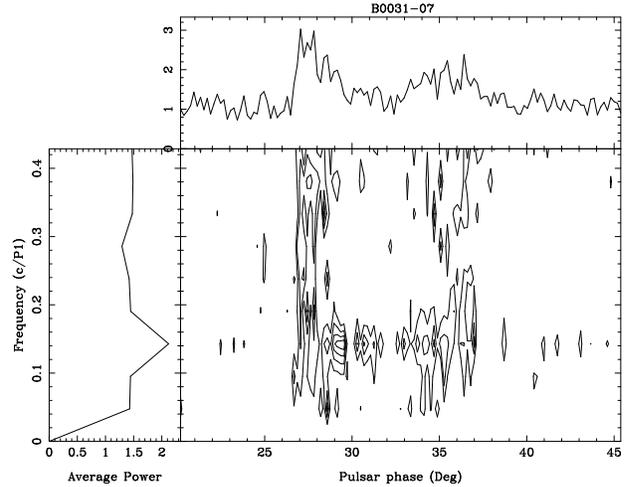}
  \caption{Contour plot of the power spectrum of the flux at 4.85\,GHz
  as a function of pulse phase during a sequence with 6-second
  periodicity. The left panel shows the power spectrum integrated over
  longitude. The upper panel shows the power integrated over
  frequency. The 0\,Hz peak has been put to 0.}
  \label{fig:FT6s}
\end{figure}

The PAPS of all three observations, are shown in
Fig.~\ref{fig:powers}. The low frequencies contain a signal due to the
nulling. To make the figure clearer we have set them to zero. We see
here that the 6-second periodicity is clearly present at 328-MHz and
is just visible in the 1.41-GHz observation. At 4.85\,GHz the PAPS
does not show the 6-second periodicity. Thus, the mode-B drift gets
weaker with increasing observing frequency. However, we did find small
sequences in the 4.85-GHz observation where there is a weak 6-second
periodicity. Fig.~\ref{fig:FT6s} shows the power spectrum of the flux
as a function of pulse phase as well as the PAPS of a sequence of 20
pulses from the 4.85-GHz observation containing 6-second
periodicity. The PAPS peaks at 6.6 seconds. We did not classify this
as a mode-B drift, because the drift-bands are not clearly visible and
the cross-correlation between consecutive pulses suggests the drift to
be in the opposite direction of all the other drifts. It would be most
difficult to explain a mode-B drift-band at high frequency that has a
drift-direction different from the drift-direction of the same mode-B
drift-band at low frequency. The present sequence is not significant
enough to establish that this has occurred.

Table~\ref{tab:P2} shows the average values for $P_3$, phase drift and $P_2$ for
each drift-mode at three frequencies.

\begin{table*}
 \flushleft
 \caption{List of average values for $P_3$ and $P_2$ for different
 drift-modes at three frequencies. The 328\,MHz and 4.85\,GHz observations were
 taken simultaneously. The values of $P_2$ have been derived from
 those of $P_3$ and $\Delta\phi$.
 }
 \begin{tabular}{cccccc}
 \hline
 \hline
  Frequency & Drift-mode & Number of & $P_3$ ($P_1$) & $\Delta\phi$ ($\degr$/$P_1$) & $P_2$ ($\degr$)\\
	    &             & Sequences &           &             &  \\
  \hline
  328\,MHz & A ($P_3\approx12\,s$)& 3 & 13.1 $\pm$ 1.1 & 1.3 $\pm$ 0.3  & 17.3 $\pm$ 1.8 \\ 
  328\,MHz & B ($P_3\approx6\,s$)& 26 & 6.8 $\pm$ 0.4 & 2.4 $\pm$ 0.3  & 16.6 $\pm$ 1.4 \\ 
  328\,MHz & C ($P_3\approx4\,s$)& 1 & 3.9 $\pm$ 0.5  & 3.54 $\pm$ 0.02 & 14 $\pm$ 1.5 \\ 
  1.41\,GHz & A ($P_3\approx12\,s$)& 7 & 11.9 $\pm$ 1.0 & 0.84 $\pm$ 0.11 & 10.0 $\pm$ 1.6 \\
  4.85\,GHz & A ($P_3\approx12\,s$)& 3 & 13.1 $\pm$ 0.5 & 0.46 $\pm$ 0.10 & 6.0 $\pm$ 1.1 \\   
 \hline
 \end{tabular}
 \label{tab:P2}
\end{table*}

Fig.~\ref{fig:polarisation} shows the average-polarisation properties
of pulses which show the same modes of drift. Each panel shows the
total intensity (solid line), linear polarisation (dashed line),
circular polarisation (dotted line) and position angle (lower half of
each panel). The left panels show the 328-MHz profiles, the right
panels show the 4.85-GHz profiles. The top panels show the average
polarisation of pulses containing subpulses with mode-A drift, below
that are the average polarisation of pulses containing subpulses with
mode-B drift and the bottom panels show the average polarisation of
all pulses. There is a clear 90$\degr$ jump in the position angles of
all pulses in both the 328-MHz and 4.85-GHz profiles at a longitude of
24$\degr$. This jump can also be seen in the pulses that only show a
mode-B drift. The widths of the average-intensity profiles are listed in
Table~\ref{tab:Widths}.

\begin{table*}[htb]
 \caption{Widths of the average intensity profiles for different
 selections of pulses from the 328-MHz, 1.41-GHz and 4.85-GHz
 observations. The 1.41\,GHz observation was not simultaneous with the
 other observations.}
 \begin{tabular}{cccc}
   \hline
   \hline
   Frequency & Which pulses & 10\% width (deg)& 50\% width (deg) \\
   \hline
   328\,MHz & Pulses in mode A ($P_3\approx12\,s$) & 37   $\pm$ 2   & 22.6 $\pm$ 1.1 \\
   328\,MHz & Pulses in mode B ($P_3\approx6\,s$) & 42.2 $\pm$ 1.1 & 22.7 $\pm$ 0.4 \\
   328\,MHz & All pulses & 39.8 $\pm$ 1.1 & 24.0 $\pm$ 0.5 \\
   1.41\,GHz & Pulses in mode A ($P_3\approx12\,s$) & 21.3 $\pm$ 1.8 & 8.8 $\pm$ 0.3 \\
   1.41\,GHz & All pulses & 29.9 $\pm$ 1.3 & 12.1 $\pm$ 0.5\\
   4.85\,GHz & Pulses in mode A ($P_3\approx12\,s$) & 24.6 $\pm$ 0.8 & 11.2 $\pm$ 0.3 \\
   4.85\,GHz & Pulses in mode B ($P_3\approx6\,s$)  & 33.5 $\pm$ 1.6 & 19.2 $\pm$ 0.5 \\
   4.85\,GHz & All pulses & 32.9 $\pm$ 0.9 & 16.1 $\pm$ 0.4 \\
   \hline
 \end{tabular}
 \label{tab:Widths}
\end{table*}

\section{Discussion}
\label{sec:discussion}
\begin{figure*}
  \centering
  \includegraphics[angle=-90, width=0.9\textwidth]{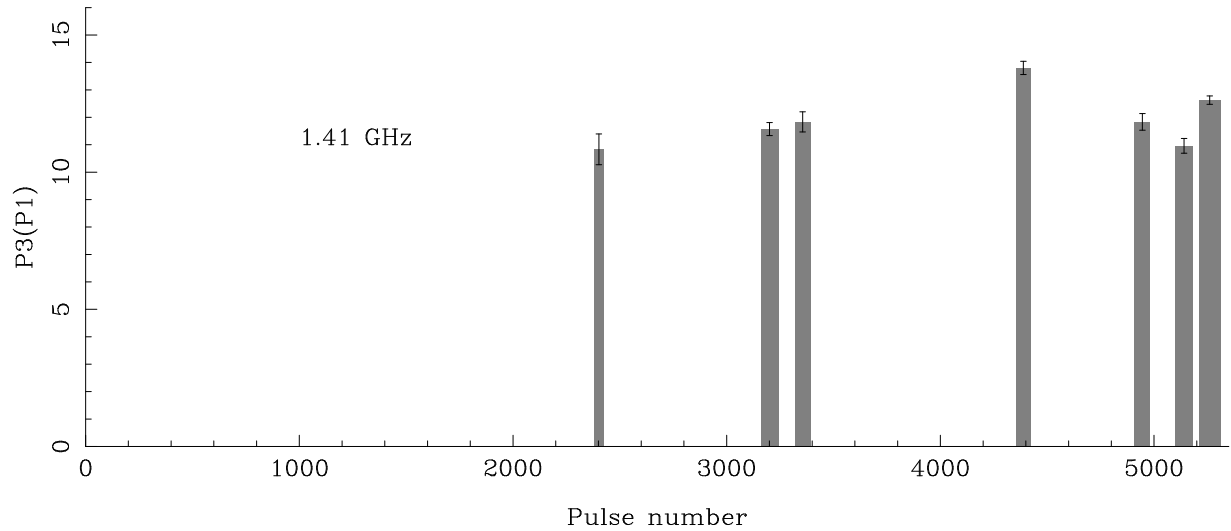}
  \caption{Sequences from the observation at 1.41\,GHz which show a
    periodicity of magnitude $P_3$ at fixed longitude. The 1.41\,GHz
    observation was not simultaneous with the other observations.}
  \label{fig:P3.1410}
\end{figure*}

We have analysed periodicities in two observations with 2700 pulses of
PSR B0031$-$07 which were taken simultaneous at 328\,MHz and
4.85\,GHz. At low frequency we found that 61.8\% of the time the
pulsar was in one of the three drift-modes. The occurrence rate was
17.8\% for mode A, 80.1\% for mode B and 2.1\% for mode C. This is
consistent with previous results of \citet{Vivekanand97}. We have
shown that whenever the mode-A drift is active, it is visible at both
frequencies. Also, when the mode-B drift is active, it is clearly
visible at 328\,MHz, but not at 1.41 or 4.85\,GHz (see
Figs.~\ref{fig:P3} and~\ref{fig:P3.1410}). However, there is 6 second
periodicity in the pulses at 1.41\,GHz and a hint of 6 second periodicity
in the pulses at 4.85\,GHz, the latter of which is possibly drifting in the
opposite direction of the drift observed at 328\,MHz. This would
suggest that towards higher frequency we are seeing less of the
drifting subpulses and begin to see a diffuse component that is also
subject to the $E\times B$ drift. It is difficult to explain a
change in the direction of drift towards higher frequency while $P_3$
remains almost constant. It might indicate that the observed drift-rate
is in fact an alias of the true drift-rate. Establishing and further
investigating the possibility of a mode-B drift at 4.85\,GHz with a
drift direction opposite to that at 328\,MHz might help determine the
actual drift-rate and direction of the subpulses of this pulsar. The
result that only one drift-mode is visible around 1.41\,GHz differs
from the results in \citet{Wright81.3} and possibly differs from
\citet{Izvekova93}.

In our observations we see that the drift-rate can change within one or two
pulses, however we do not see any instance of a mode change within a
drift-band. It should be noted that in most cases there is at least one
null between two drift-bands with different drift-rates.

Table~\ref{tab:P2} shows that the $P_2$ of drift-modes A~and~B at
328\,MHz is almost the same and that drift-mode C has a slightly
smaller $P_2$ at this frequency. Within errors, however, our values
agree with a constant $P_2$ for each drift-mode, unlike the behaviour
predicted by \citet{Vivekanand97}, who claim that $P_2$ increases
monotonically with $\Delta\phi$. With increasing frequency, the value
of $P_2$ for polarisation mode A decreases, which can be explained by
the decrease of the opening angle towards higher frequency, assuming
radius-to-frequency mapping.

\begin{figure*}[ht!]
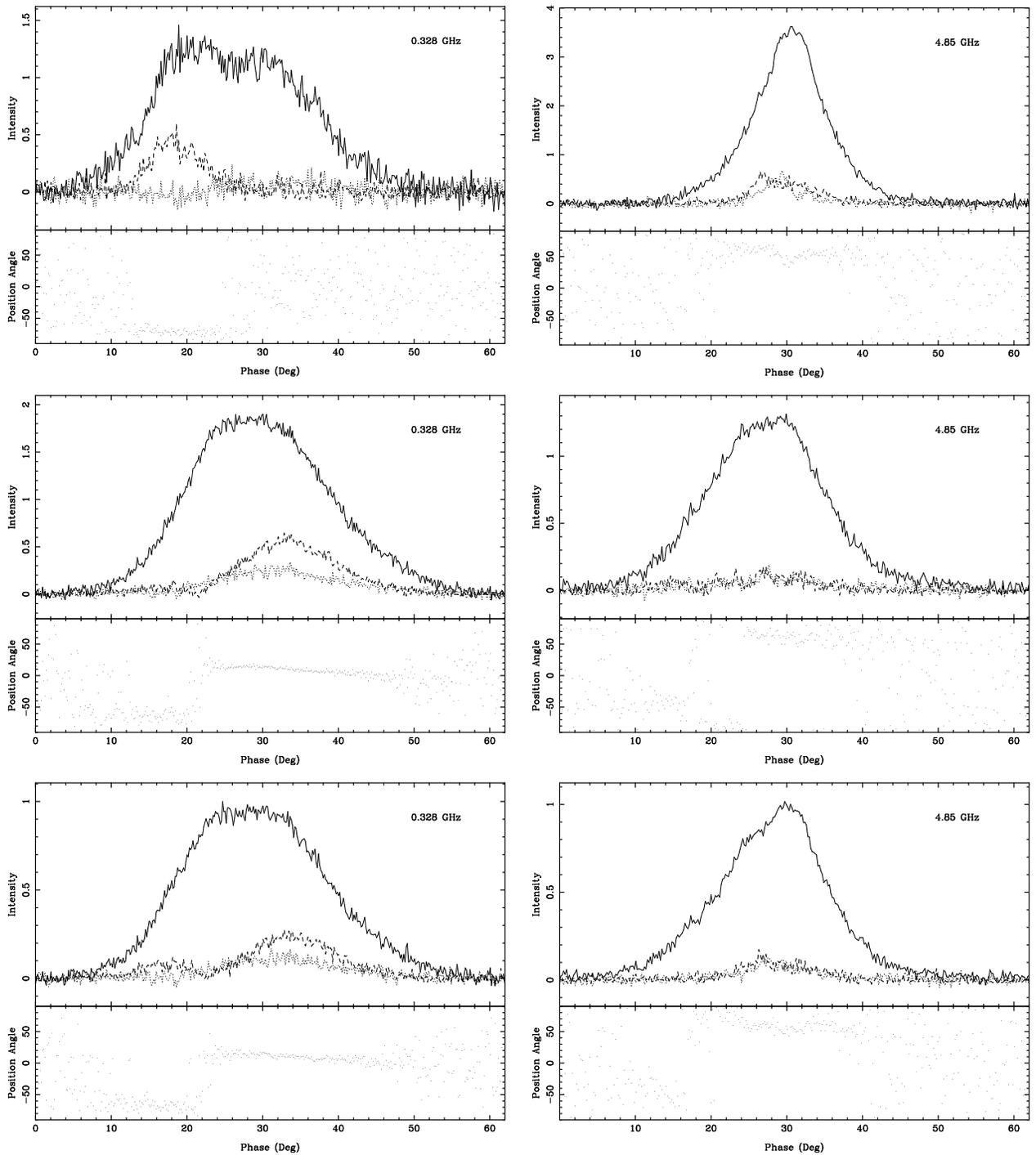

  \centering
  \begin{tabular}{cc}
  \includegraphics[angle=-90, width=0.45\textwidth]{1626f11.eps} &
  \includegraphics[angle=-90, width=0.45\textwidth]{1626f12.eps} \\
  \includegraphics[angle=-90, width=0.45\textwidth]{1626f13.eps} &
  \includegraphics[angle=-90, width=0.45\textwidth]{1626f14.eps} \\
  \includegraphics[angle=-90, width=0.45\textwidth]{1626f15.eps} &
  \includegraphics[angle=-90, width=0.45\textwidth]{1626f16.eps} \\
  \end{tabular}
  \caption{Average intensity and polarisation of pulses in mode A (top
    panels), pulses in mode B (middle panels) and all pulses (lower
    panels). The profiles have been normalised to set the peak in the
    average intensity of all pulses to 1. The left panels show the
    average of pulses at 328\,MHz, the right panels show the average
    of the same pulses at 4.85\,GHz. Since the pulses at
    4.85\,GHz do not show a mode-B drift, the right middle panel
    contains the average of pulses at 4.85\,GHz that show a mode-B
    drift at 328\,MHz. The solid, dashed and dotted lines in each
    panel represent the total intensity, linear polarisation and
    circular polarisation, respectively. The lower part of the panels
    show the polarisation position angle. The 328\,MHz profiles have
    been re-binned to 500\,$\mu$s per bin, to make the profiles
    comparable.}
\label{fig:polarisation}
\end{figure*}

Fig.~\ref{fig:polarisation} shows that there is a distinct difference
between the average profile of pulses with a 12-second periodicity
(A-profile) and with a 6-second periodicity (B-profile). At 328\,MHz,
the A-profile seems to have two components. This can correspond to the
line of sight cutting the edge of the subpulses in the centre of the
profile, thereby bifurcating the average profile.  It is interesting
that the right component in the A-profile at 328\,MHz seems to
correspond to the single component of the A-profile at 4.85\,GHz. Thus
it appears as though the first component in the A-profile at 328\,MHz
disappears towards higher frequency. The profiles at 328\,MHz also
show that the intensity of the pulses in drift-mode A is on average
lower than the intensity of the pulses in drift-mode B. A difference
in average profile of the three modes has been reported before by
\citet{Wright81.3} and \citet{Vivekanand97}. The former authors have
observed the pulsar at 1.62\,GHz and found that the A-profile is more
narrow than the B-profile, which is in turn more narrow than the
C-profile. We cannot directly compare this result with our observation
at 4.85\,GHz as we do not see a mode-B drift at this frequency. But if
we define the B-profile as the pulses at 4.85\,GHz that show a mode-B
drift at 328\,MHz, then we find that at 4.85\,GHz, the A-profile is
indeed more narrow than the B-profile. \citet{Wright81.3} do not note
a difference in amplitude, nor an offset which are first noted by
\citet{Vivekanand97}, who have observed the pulsar at 326.5\,MHz. They
show that at this frequency pulses in drift-mode~B have on average
more intensity than pulses in drift-mode A~and~C, which are of equal
intensity. They also state that drift-mode~A arrives earlier than
drift-mode~B, which in turn arrives earlier than drift-mode~C. Both
findings are confirmed by our results from the 328\,MHz
observation.  However, they do not report a double component in the
A-profile, which is indeed not present in their plot. This might
be due to the fact that only a single linear polarisation was used in
their observation. Table~\ref{tab:Widths} shows that the widths from
the average intensity profiles decreases from 328\,MHz to 1.41\,GHz
and increases again from 1.41 to 4.85\,GHz. This behaviour is not
consistent with radius-to-frequency mapping. However, we should note
that the signal-to-noise ratio of the edges of the profiles might not
be high enough to detect the entire width of the pulse, which makes it
difficult to draw any conclusions from these
values. Furthermore, when we compare the change in 50\%-width
from the A-profile with the change in $P_2$ between 328\,MHz and
4.85\,GHz, we find that the value for $P_2$ decreases roughly by 65\%
while the width of the A-profile only decreases by 50\%. Both should
reflect the change in the size of the radioactive region due to
radius-to-frequency mapping. To explain this discrepancy we suggest
that at each frequency the drift-path of the subbeams is surrounded by
an area of weak radio-emission which becomes relatively smaller with
decreasing frequency. To support this claim we have constructed
average intensity profiles with contributions from the mode-A
drift-bands only. This was achieved by adding up the spectral power in
the fluctuation spectrum in the domain between 0.07 and 0.1 Hz at each
pulse phase. We then compared the 50\%-widths of these profiles at
328\,MHz and 4.85\,GHz. We found that these widths were approximately
22$\degr$ and 7$\degr$, respectively, which is consistent with the
change of $P_2$ between these frequencies.

Fig.~\ref{fig:polarisation} also shows that the average polarisation
of all pulses from the 328-MHz observation has two components and a
clear minimum around a pulsar phase of 24$\degr$. From the average
polarisation of pulses at 328\,MHz that are in drift-mode A and B, it
is apparent that the pulses in drift-mode A contribute only to the
component on the left and the pulses in drift-mode B contribute only
to the component on the right. The average position angle of all
pulses shows a 90$\degr$ jump at both frequencies at a pulsar phase of
24$\degr$. This can be interpreted as two orthogonally polarised modes
changing dominance at this pulsar phase. A straight line fit to the
position angles of both modes has shown that these polarisation modes
are indeed 90$\degr$ apart. This jump is also visible in the average
position angle of pulses that are in mode-B drift. We have searched
for this jump in the average position angle of short sequences
containing 20 to 50 pulses in a particular drift-mode and did not find
a clear change in polarisation mode. It only manifests itself in the
average of many sequences. Furthermore, the pulses in mode-A drift do
not show any sign of a jump in position angle. They only show some
degree of linear polarisation in the left part of the profile, just
before the mode jump occurs in the average position angle of all
pulses. Thus the left part of the A-profile is dominated by one
of the two orthogonal polarisation modes, while the lack of
polarisation in the right part of the A-profile suggests that here
both polarisation modes are of equal strength. The lack of
polarisation in the left part of the B-profile suggests that here both
polarisation modes are also of equal strength, while the right part of
the B-profile is dominated by the other polarisation mode. This means
that there is a strong relationship between the drift-modes A and B
and the two orthogonal modes of polarisation.

\begin{figure*}
  \centering
  \begin{tabular}{cc}
  \includegraphics[width=0.5\textwidth]{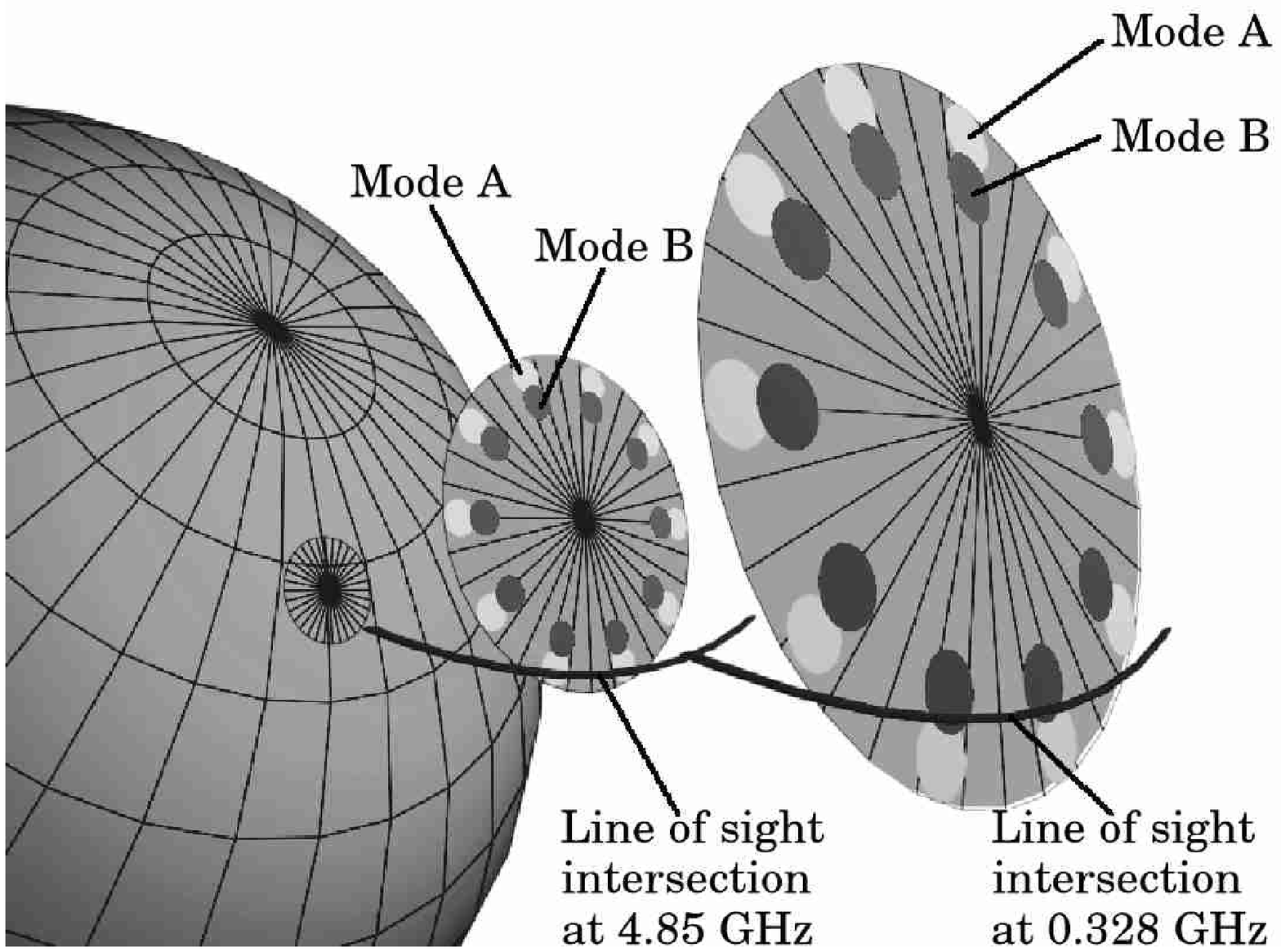} 
&
  \includegraphics[width=0.5\textwidth]{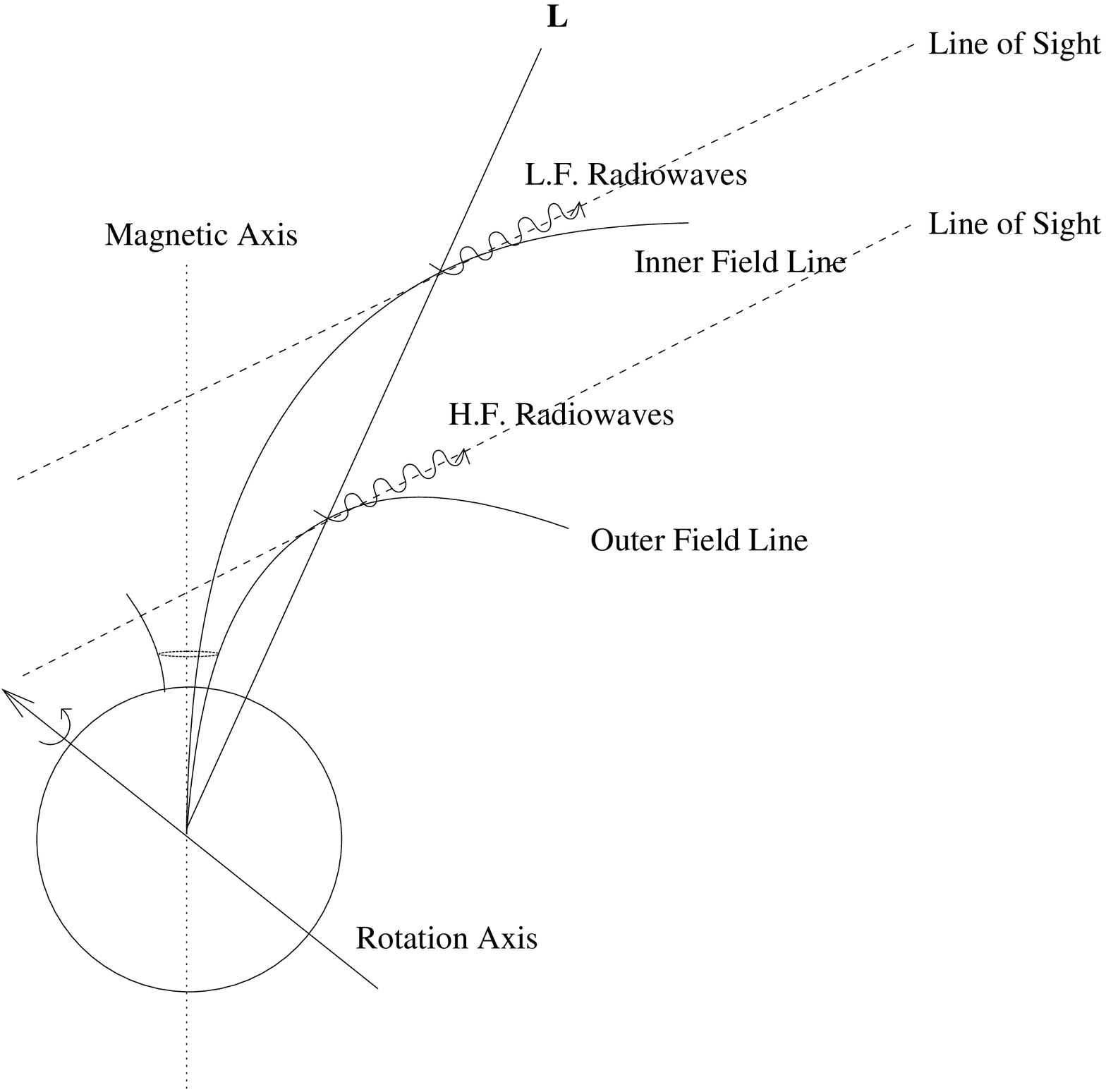} 
  \\
  (a) & (b) \\
  \end{tabular}
 \caption{(a) Schematic overview of the proposed geometrical model to
  explain the absence of one mode at high frequency. The two large
  discs are centred around the magnetic axis and represent the
  emission region at two different frequencies, corresponding to two
  different altitudes above the pulsar surface. The smaller circles in
  the emission region represent the positions of the drifting
  subpulses, which rotate around the magnetic axis. The true number of
  subpulses is unknown. The different drift-modes are illustrated by
  different colours. Please note that only one drift-mode is assumed
  to be active at a time. (b) Schematics of two field lines from which
  radio waves of two different frequencies are observed. L is the line
  that connects the locations where the field lines are directed
  towards the observer. Due to radius-to-frequency mapping, low
  frequency radio waves are emitted from inner field lines with
  respect to those where the high frequency radiation is emitted for
  the same line of sight. This picture also illustrates that radio
  waves observed from outer field lines are emitted closer to the
  magnetic axis than radio waves from inner field lines. Thus, the
  width of the emission profile decreases with increasing frequency.}
 \label{fig:pulsar}
\end{figure*}

\subsection {Modelling the observations}
Since the limited length of our observations does not enable us to
study mode C, we shall only attempt to model the behaviour of modes A
and B, which are the most prominent drift-modes. We have found that of
these two drift-modes only mode A is visible at high frequency, while
at low frequency both modes are visible and are seen to have different
orthogonal polarisation. At low frequency, the pulses in mode A have
less intensity than the pulses in mode B, while at high frequency the
pulses show the opposite behaviour. Furthermore, even though the mode
B drift is not clearly visible at 4.85\,GHz, we do see a hint of 6
seconds periodicity at this frequency. In the context of the potential
gap model, the different rates of drift seem to suggest that the
potential gap can take on different stable values. Each stable value
can be associated with a particular drift rate and a particular
magnetic surface of emission if we assume that the radiation is
emitted tangential to the magnetic field lines. An increase in the
value of the potential gap is expected to give rise to a faster drift
and emission (pair production) from magnetic field-lines closer to the
magnetic axis. This leads to the picture of an emission region with
two concentric radiating rings. At high frequency, the line of sight
intersects with magnetic field lines which are further away from the
magnetic axis than at low frequency. Therefore, the drifting component
of the inner ring is not seen at high frequency. This is graphically
illustrated in Fig.~\ref{fig:pulsar}.

\begin{figure*}
  \centering
  \begin{tabular}{cc}
  \includegraphics[width=0.5\textwidth]{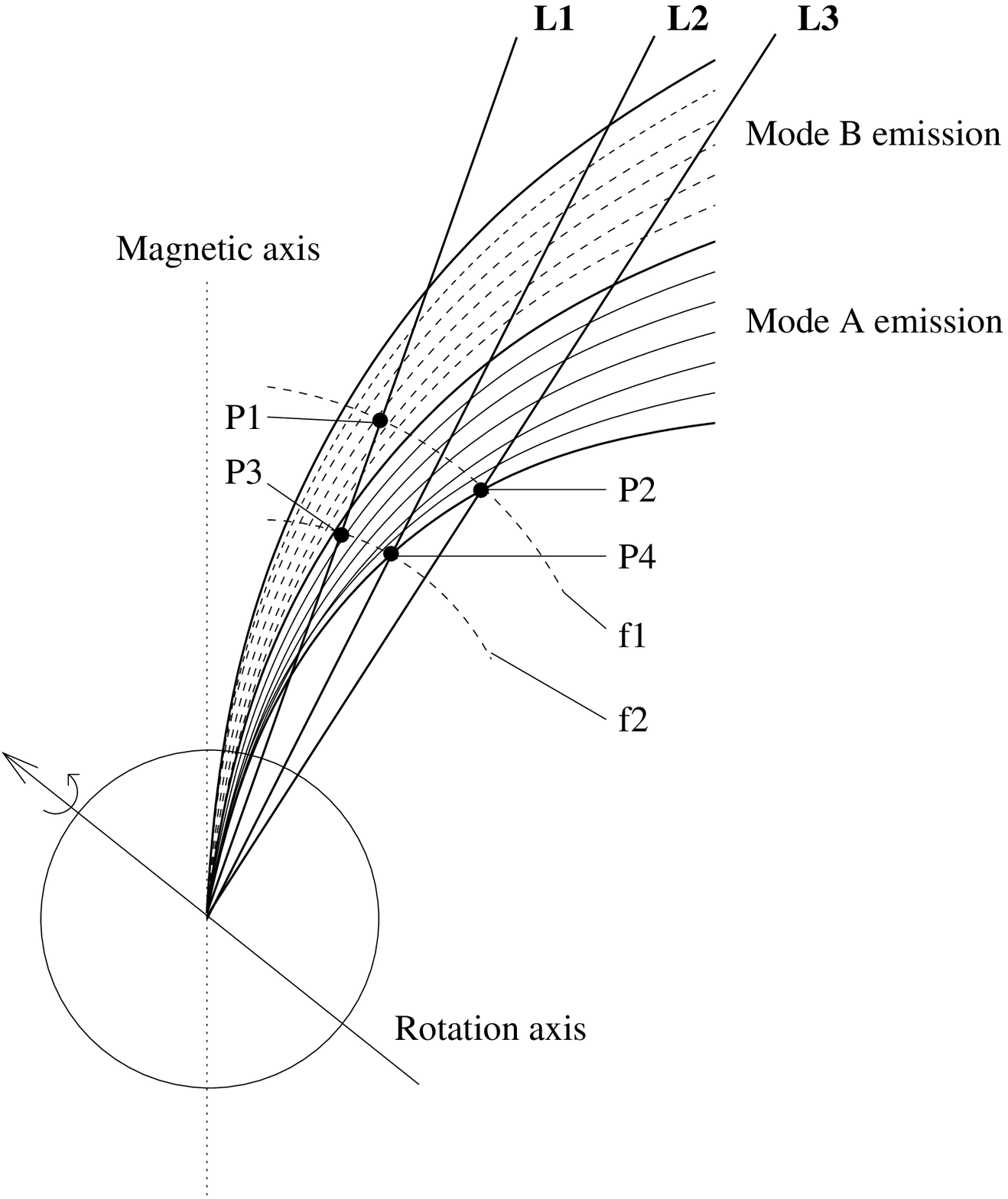} 
  &
  \includegraphics[width=0.435\textwidth]{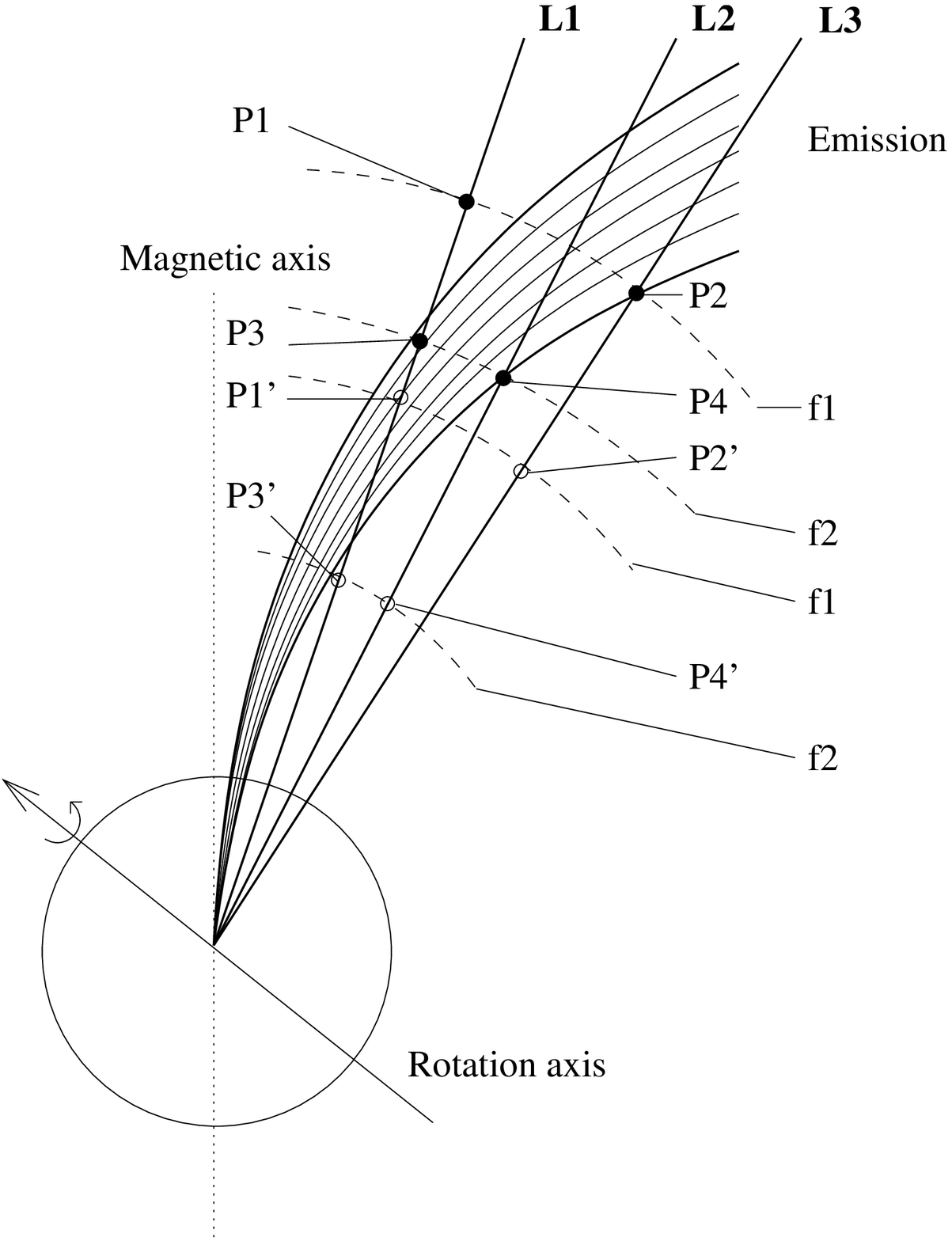} 
  \\
  \end{tabular}
 \caption{Schematic overview of two models explaining the observed
 behaviour of two of the drift-modes of PSR B0031$-$07 at two
 frequencies f1, f2 (f1$<$f2). The left panel illustrates our present model. The
 right panel illustrates the model of \citet{Leeuwen03}. As the
 pulsar rotates the line of sight passes from L3 over L2 to L1 and
 back. Emission at low frequency corresponds to altitudes P1, P2 (and
 P1', P2') whereas emission at high frequency comes from altitudes P3,
 P4 (and P3', P4').}
 \label{fig:models}
\end{figure*}

It is also possible to construct a model wherein the emission of both
modes comes from the same magnetic flux surface, but from different
heights. Such a model is given in \citet{Leeuwen03}. The difference
between the two models is graphically illustrated in
Fig.~\ref{fig:models}. Here, L1, L2 and L3 are the locations where the
field lines are directed towards the observer for three different
lines of sight: L1 when the line of sight is closest to the magnetic
axis, L2 when the line of sight just touches the emission region when
observing at high frequency, and L3 shows when the line of sight just
touches the emission region when observing at low frequency. In case
of a dipole magnetic field, the lines L1 - L3 are straight. Note that
these lines do not lie in a plane as might be suggested by
Fig.~\ref{fig:models}. Therefore, one must consider these images as
lying in the plane through one of the three lines and the magnetic
axis. At points on L1, L2 or L3 further away from the pulsar surface
the curvature of the field lines decreases and the frequency of
emission is believed to be lower.

In our model (left image) a mode change is caused by a shift of the
emission to inner field-lines. At high frequency, the observer can only
see emission coming from a region between points P3 and P4. As the
mode changes from A to B, the observer will no longer see emission. At
low frequency the observer can only see emission coming from a region
between the points P1 and P2, and is thus able to see both modes. Note
that when the line of sight is closest to the magnetic axis, mode A is
not visible, thus causing a dip in the centre of the A-profile at low
frequency. 

In the model of \citet{Leeuwen03} (right image) the emission always
comes from the same field-lines. Now however, the emission  altitude
at a fixed frequency decreases when the mode changes from A to
B. This causes the observer to see emission from a point lower in the
pulsar magnetosphere (indicated by an open dot and an apostrophe). At
high frequency, the observer will only see emission coming from a
region between P3 and P4 if the pulsar is emitting in mode A but none
in mode B since the region between P3' and P4' is not part of the
'active' (radiating) flux tube. Thus the observer will only see emission in mode
A. At low frequency the observer will only see emission coming from a
region between P1 and P2 if the pulsar is emitting in mode A and will
only see emission in a region between P1' and P2' if the pulsar is
emitting in mode B. Thus the observer will be able to see emission
from both modes. Note that also in this model mode A is not visible at
low frequency when the line of sight is closest to the magnetic
axis. 

In the case of PSR B0031$-$07 both models can explain the observed
characteristics. It is therefore difficult to distinguish between them
observationally, especially if observing at only one
frequency. However, if we were to observe this pulsar at a large range
of frequencies (preferably simultaneous) it should be possible to make
quantitative statements as to favour one of the models. 

On theoretical grounds, we find a possible inconsistency in the model
of \citet{Leeuwen03} when applied to our observations. In their model,
which follows the work of \citet{Ruderman75} and \citet{Melikidze00},
the transition from mode A to B occurs by a decrease in height of the
voltage gap. As a result the altitude of emission decreases. However,
the electric field also decreases and with it the speed of the
$E\times B$ drift, contrary to what is observed. Of course, a possible
explanation for this inconsistency could be that we are not seeing the
actual drift-speed, but an alias. It should be noted that variation of
drift speed can also happen due to temperature variation on the polar
cap as suggested by \citet{Gil03}.

\section{Conclusions}
From an analysis of 2700 pulses from PSR B0031$-$07 taken
simultaneously at 328\,MHz and 4.85\,GHz we found that from the three
known drift-modes A, B and C of PSR B0031$-$07 only mode A is visible
at high frequencies. We have constructed a geometrical model that
explains how one drift-mode can disappear at high frequency while
another drift-mode remains visible. Further, we have shown that the two
most prominent drift-modes A and B are associated with two orthogonal
modes of polarisation, respectively. To continue the study
presented here, one would require more multi-frequency single pulse
observations from this pulsar containing full Stokes parameters.

\begin{acknowledgements}
The authors would like to thank J. Gil for his suggestions and
discussion towards interpretation of the results and A. Jessner,
A. Karastergiou, B. Stappers and our referee, J. Rankin, for their
helpful discussions. We also thank all the members of the MFO
collaboration for the establishment of the project which led to the
observations being available. This paper is based on observations with
the 100-m telescope of the MPIfR (Max-Planck-Institut f\"ur
Radioastronomie) at Effelsberg and the Westerbork Synthesis Radio
Telescope (WSRT) and we would like to thank the technical staff and
scientists who have been responsible for making these observation
possible.
\end{acknowledgements}

\bibliographystyle{aa}
\bibliography{1626}

\begin{thebibliography}{22}
\expandafter\ifx\csname natexlab\endcsname\relax\def\natexlab#1{#1}\fi

\bibitem[{{Deshpande} \& {Rankin}(1999)}]{Deshpande99}
{Deshpande}, A.~A. \& {Rankin}, J.~M. 1999, \apj, 524, 1008

\bibitem[{{Drake} \& {Craft}(1968)}]{Drake68}
{Drake}, F.~D. \& {Craft}, H.~D. 1968, Nature, 220, 231

\bibitem[{{Edwards} \& {Stappers}(2004)}]{Edwards04}
{Edwards}, R.~T. \& {Stappers}, B.~W. 2004, \aap, 421, 681

\bibitem[{{Gil} {et~al.}(2003){Gil}, {Melikidze}, \& {Geppert}}]{Gil03}
{Gil}, J., {Melikidze}, G.~I., \& {Geppert}, U. 2003, \aap, 407, 315

\bibitem[{{Huguenin} {et~al.}(1970){Huguenin}, {Taylor}, \&
  {Troland}}]{Huguenin70}
{Huguenin}, G.~R., {Taylor}, J.~H., \& {Troland}, T.~H. 1970, \apj, 162, 727

\bibitem[{{Izvekova} {et~al.}(1993){Izvekova}, {Kuzmin}, {Lyne}, {Shitov}, \&
  {Smith}}]{Izvekova93}
{Izvekova}, V.~A., {Kuzmin}, A.~D., {Lyne}, A.~G., {Shitov}, Y.~P., \& {Smith},
  F.~G. 1993, \mnras, 261, 865

\bibitem[{{Joshi} \& {Vivekanand}(2000)}]{Joshi00}
{Joshi}, B.~C. \& {Vivekanand}, M. 2000, \mnras, 316, 716

\bibitem[{{Krishnamohan}(1980)}]{Krishnamohan80}
{Krishnamohan}, S. 1980, \mnras, 191, 237

\bibitem[{{Kuzmin} {et~al.}(1986){Kuzmin}, {Malofeev}, {Izvekova}, {Sieber}, \&
  {Wielebinski}}]{Kuzmin86}
{Kuzmin}, A.~D., {Malofeev}, V.~M., {Izvekova}, V.~A., {Sieber}, W., \&
  {Wielebinski}, R. 1986, \aap, 161, 183

\bibitem[{{Manchester} {et~al.}(1975){Manchester}, {Taylor}, \&
  {Huguenin}}]{Manchester75}
{Manchester}, R.~N., {Taylor}, J.~H., \& {Huguenin}, G.~R. 1975, \apj, 196, 83

\bibitem[{{Melikidze} {et~al.}(2000){Melikidze}, {Gil}, \&
  {Pataraya}}]{Melikidze00}
{Melikidze}, G.~I., {Gil}, J.~A., \& {Pataraya}, A.~D. 2000, \apj, 544, 1081

\bibitem[{{Radhakrishnan} \& {Cooke}(1969)}]{Rad69}
{Radhakrishnan}, V. \& {Cooke}, D.~J. 1969, \aplett, 3, 225

\bibitem[{{Rankin}(1986)}]{Rankin86}
{Rankin}, J.~M. 1986, \apj, 301, 901

\bibitem[{{Ruderman} \& {Sutherland}(1975)}]{Ruderman75}
{Ruderman}, M.~A. \& {Sutherland}, P.~G. 1975, \apj, 196, 51

\bibitem[{{Taylor} {et~al.}(1993){Taylor}, {Manchester}, \& {Lyne}}]{Taylor93}
{Taylor}, J.~H., {Manchester}, R.~N., \& {Lyne}, A.~G. 1993, \apjs, 88, 529

\bibitem[{{van Leeuwen} {et~al.}(2002){van Leeuwen}, {Kouwenhoven},
  {Ramachandran}, {Rankin}, \& {Stappers}}]{Leeuwen02}
{van Leeuwen}, A.~G.~J., {Kouwenhoven}, M.~L.~A., {Ramachandran}, R., {Rankin},
  J.~M., \& {Stappers}, B.~W. 2002, \aap, 387, 169

\bibitem[{{van Leeuwen} {et~al.}(2003){van Leeuwen}, {Stappers},
  {Ramachandran}, \& {Rankin}}]{Leeuwen03}
{van Leeuwen}, A.~G.~J., {Stappers}, B.~W., {Ramachandran}, R., \& {Rankin},
  J.~M. 2003, \aap, 399, 223

\bibitem[{{Vivekanand}(1995)}]{Vivekanand95}
{Vivekanand}, M. 1995, \mnras, 274, 785

\bibitem[{{Vivekanand} \& {Joshi}(1997)}]{Vivekanand97}
{Vivekanand}, M. \& {Joshi}, B.~C. 1997, \apj, 477, 431

\bibitem[{{Vivekanand} \& {Joshi}(1999)}]{Vivekanand99}
{Vivekanand}, M. \& {Joshi}, B.~C. 1999, \apj, 515, 398

\bibitem[{{Wright}(1981)}]{Wright81}
{Wright}, G.~A.~E. 1981, \mnras, 196, 153

\bibitem[{{Wright} \& {Fowler}(1981)}]{Wright81.3}
{Wright}, G.~A.~E. \& {Fowler}, L.~A. 1981, in IAU Symp. 95: Pulsars: 13 Years
  of Research on Neutron Stars, 211--+

\end{thebibliography}

\clearpage

\clearpage

\clearpage
\end{document}